\documentclass[12pt]{article} 
\usepackage{amssymb,amsmath,color,graphicx,float}

\numberwithin{equation}{section}
\usepackage[toc,page]{appendix}

\newcommand{\nc}{\newcommand}

\nc{\la}{\lambda} \nc{\alf}{\alpha} \nc{\La}{\Lambda} \nc{\ze}{\zeta}
\nc{\tht}{\theta} \nc{\T}{\Theta} \nc{\be}{\beta}  \nc{\eps}{\epsilon} 
\nc{\ga}{\gamma}  \nc{\De}{\Delta}  \nc{\G}{\Gamma}  \nc{\vphi}{\varphi}
\nc{\de}{\delta} \nc{\si}{\sigma}  \nc{\ka}{\kappa}   \nc{\Si}{\Sigma} 
\nc{\om}{\omega}  \nc{\qq}{\qquad}                \nc{\Om}{\Omega} \nc{\vrho}{\varrho}
\nc{\nf}{\infty}   \nc{\dl}{\mathop{\smash{\cal L}}}  \nc{\black}{\rule{3mm}{3mm}}
\nc{\ra}{\rightarrow}    \nc{\ol}{\overline}        \nc{\und}{\underline} 
\nc{\beq}{\begin{equation}}  \nc{\eeq}{\end{equation}}  \nc{\pt}{\partial}  
   \nc{\dst}{\displaystyle}  \nc{\na}{\nabla} 
\nc{\nnb}{\nonumber}    \nc{\bs}{\backslash}        \nc{\mb}{\mathbb}   
\nc{\sn}{{\rm sn}\,} \nc{\cn}{{\rm cn}\,}     \nc{\dn}{{\rm dn}\,} \nc{\nin}{\noindent}
\nc{\ti}{\tilde}   \nc{\wti}{\widetilde}   \nc{\h}{\hat}  \nc{\wh}{\widehat}
\nc{\tpsi}{\wti{\psi}}   \nc{\tphi}{\wti{\phi}}  \nc{\tH}{\wti{H}} \nc{\Ai}{{\rm Ai}}

\nc{\Pf}{P_{\phi}}  \nc{\Pt}{P_{\tht}}
\nc{\Iph}{I_{\phi}}  \nc{\Ith}{I_{\tht}}
\nc{\omt}{\om_{\tht}}  \nc{\omp}{\om_{\phi}}

\newcounter{muni}
\newenvironment{remunerate}{\begin{list}{{\rm \arabic{muni}.}}
{\usecounter{muni}
\setlength{\leftmargin}{0pt}\setlength{\itemindent}{38pt}}}{\end{list}}

\nc{\brm}{\begin{remunerate}}   \nc{\erm}{\end{remunerate}}

 \newtheorem{nlem}{Lemma}
\newtheorem{nth}{Proposition}  \newtheorem{nTh}{Theorem}
\newtheorem{ndef}{Definition}     

\nc{\stg}{\mathop{\smash{*}}}
\nc{\st}{\mathop{\smash{\delta}}}
\nc{\barr}{\begin{array}}   \nc{\earr}{\end{array}}   \nc{\dg}{\dagger}
\nc{\mtvb}{\mathversion{bold}}   \nc{\mtvn}{\mathversion{normal}} 

\begin{document}

%\date{\today}

\begin{titlepage}

\vskip 0.5truecm\centerline{\Large\bf Superintegrable geodesic flows}

\vskip 0.5truecm 

\vskip 0.5truecm\centerline{\Large\bf on the hyperbolic plane} 

\vskip 1.0truecm
\centerline{ \bf Galliano VALENT}  

\vspace{1cm}
\centerline{LPMP: Laboratoire de Physique Math\'ematique de Provence}

\vspace{3mm} 
\centerline{13100 Aix en Provence, France.} 

\vskip 1.5truecm

\begin{abstract} In the framework laid down by Matveev and Shevchishin, superintegrability is achieved with one integral linear in the momenta (a Killing vector) and two extra integrals of of any degree above two in the momenta. However these extra integrals may exhibit either a trigonometric dependence in the Killing coordinate (a case we have already solved) or a hyperbolic dependence and this case is solved here. Unfortunately the resulting geodesic flow is {\em never} defined on the two-sphere, as  was the case for Koenigs systems (with quadratic extra integrals). Nevertheless we give some sufficient conditions under which the geodesic flow is defined on the hyperbolic plane. 
\end{abstract}

\vspace{2cm}

\noindent MSC (2010): {\tt 32C05}, {\tt 53C22}, 
{\tt 37E99}, {\tt 37J35}, {\tt  37K25}, {\tt 81V99}.

\end{titlepage}

\newpage
\section{Introduction}
Matveev and Shevchishin \cite{ms} have proposed an interesting approach to superintegrability on surfaces of revolution equipped with the hamiltonian
\beq
H=h_x^2(P_x^2+P_y^2),\qq\quad h_x=D_x h(x).
\eeq
This system is integrable and to reach superintegrability we need some extra integral, let us say $S_1$,  which implies in turn the existence of a second one, namely $S_2=\{P_y,S_1\}$.

{\em Under appropriate hypotheses}, they have shown that the 
$y$ dependence of the extra integral (which was supposed to be cubic in the momenta) may be of three kinds:
\brm
\item In the so called ``affine case" the extra integral can be quadratic:
\[ S_1=S_{1,0}+S_{1,1}y +S_{1,2}y^2.\]
\item In the second case the extra integral can be trigonometric
\[ S_1=\cos y\,{\cal S}+\sin y\,{\cal T}.\]
\item In the third case the  extra integral can be 
hyperbolic
\[S_1=\cosh y\,{\cal S}+\sinh y\,{\cal T}.\]
\erm
In fact a prior work by Koenigs \cite{Ko}, popularized and generalized in \cite{kk1} and \cite{kk2}, involving a completely different analysis, had established that for a {\em quadratic} extra integral  only the three issues stated above were possible. Unfortunately the geodesic flow of Koenigs superintegrable (SI) systems never meet ${\mb S}^2$ as emphasized in \cite{Va2}.   

We gave the solution for the affine case in \cite{Va3} by reducing the problem to a linear ODE for which the general solution could be obtained explicitly.

We solved the trigonometric case in \cite{Va4}: as foreseen by Matveev and Shevchishin in this case one runs into Zoll geometry (see \cite{Be} and \cite{Zo} for this subject) and we could give sufficient conditions insuring that the geodesic flow is globally defined on $M={\mb S}^2$. This seems to be the most interesting situation.

The remaining hyperbolic case is the subject of this article, but, as we will see, despite the superintegrability of its geodesic flow it never meets 
${\mb S}^2$. Nevertheless, some sufficient conditions do give a quite large panel of solutions globally defined on ${\mb H}^2$. 

This article has the following content: in Section 2 the main results are stated. 

Then, in a First Part, we examine the case where $\sharp(S)=2n\geq 2$, beginning with a presentation of the framework laid down by Matveev and Shevchishin in Section 3, followed by the proofs of Theorems 1 and  2  in Sections 4 and 5. In Section 6 and for quadratic extra integrals we relate our results with Koenigs ones. 

We proceed to the Second Part dealing with integrals of order $2n+1\geq 3$. In Section 7 the setting for this case is given and we prove Theorems 3 and 4 respectively in Sections 8 and 9. 

The global issues are discussed in Section 10, where Theorem 5 is proved, and some concluding remarks are 
presented in Section 11.

The article ends up with Appendices A and B in which we prove several technical relations.

%%%%%%%%%%%%%%%%%%%%%%%%%%

\section{The results}
As explained in the introduction we will consider the geodesic flow of the hamiltonian 
\beq\label{ham1}
H=\Pi^2+\frac{P_y^2}{\cosh^2 t},\qq\quad \Pi=\frac{P_t}{A(t)}.
\eeq
It does exhibit one linear integral $P_y$, and to reach superintegrability we have to construct two extra integrals
$S_1$ and $S_2$ which are polynomials in $H$ and in 
$P_y^2$, of fixed degree, denoted by 
$\sharp(S_1)=\sharp(S_2)$, in the momenta.

\subsection{Integrals of even degree}
Let us first consider the case where $\sharp(S_1)=\sharp(S_2)=2n$ with $n\geq 1$ and the set of integrals
\beq
H,\qq \Pf, \qq S_1=\cosh y\,{\cal S}+\sinh y\,{\cal T}, \qq S_2=\cosh y\,{\cal T}-\sinh y\,{\cal S},\eeq
where  
\beq
{\cal S}=\sum_{k=0}^n\,\la_{2k-1}(\tht)\,H^{n-k}\,P_y^{2k}, \qq\qq {\cal T}=\Pi\sum_{k=0}^n\,\la_{2k}(t)\,H^{n-k}\,P_y^{2k}.
\eeq  
We will prove:

\begin{nTh} The geodesic flow of the hamiltonian (\ref{ham1}) is superintegrable if one takes
\beq\label{A1}
A(t)=1+\sum_{k=1}^{2n-1}\,\frac{e_k\,\sinh t}{\sqrt{m_k\,\cosh^2 t-1}}, \qq\quad \forall k: \quad e_k=\pm 1, 
\eeq
where all of the $2n-1$ real parameters $m_k$ are restricted to $m_k>1$. \end{nTh}

\begin{nTh}
The set of integrals
\[ H,\qq P_y, \qq S_{\pm}=S_1\pm S_2,  \]
generates a Poisson algebra, with the relations 
\beq
S_+\,S_-=\sum_{k=0}^{2n}\,\si_k\,H^{2n-k}\,P_y^{2k}, \qq 
\{S_+,S_-\}=-2\sum_{k=0}^{2n-1}\,(k+1)\si_{k+1}\,H^{2n-k}\,P_y^{2k-1},\eeq
and the coefficients $\si_l$ are explicitly given in terms of the coefficients $m_k$, see (\ref{Simom}).\end{nTh}

\subsection{Integrals of odd degree}
In the case where $\sharp(S_1)=\sharp(S_2)=2n+1$ with $n\geq 1$, we will consider the set of integrals
\beq
H,\qq P_y, \qq S_1=\cosh y\,{\cal S}+\sinh y\,{\cal T}, \qq S_2=\sinh y\,{\cal S}+\cosh y\,{\cal T},
\eeq
where
\beq\label{AB2}
{\cal S}=\Pi\sum_{k=0}^n\,\la_{2k-1}(\tht)\,H^{n-k}\,P_y^{2k}, \qq\qq {\cal T}=\sum_{k=0}^{n-1}\,\la_{2k}(\tht)\,H^{n-k-1}\,P_y^{2k+1},
\eeq  
and we get similar results:
\begin{nTh} The geodesic flow of the hamiltonian (\ref{ham1}) is superintegrable if one takes
\beq\label{A1}
A(t)=1+\sum_{k=1}^{2n}\,\frac{e_k\,\sinh t}{\sqrt{m_k\,\cosh^2 t-1}}, \qq\quad \forall k: \quad e_k=\pm 1, 
\eeq
where all of the $2n$ real parameters $m_k$ are restricted to $m_k>1$. \end{nTh}

\begin{nTh}
The set of integrals
\[ H,\qq P_y, \qq S_{\pm}=S_1\pm S_2,  \]
generates a Poisson algebra, with the relations 
\beq
S_+\,S_-=\sum_{l=0}^{2n}\,\si_l\,H^{2n-l}\,P_y^{2l}, \qq 
\{S_+,S_-\}=-2\sum_{l=0}^{2n-1}\,(l+1)\si_{l+1}\,H^{2n-1-l}\,P_y^{2l+1},\eeq
and the coefficients $\si_l$ are explicitly given in terms of the coefficients $m_k$.
\end{nTh}

\subsection{The global issues} 
They strongly depend on the degree of the integrals. We have:

\begin{nTh} The superintegrable geodesic flows considered in this article:
\begin{itemize}
\item[a)] Are {\bf never} globally defined on ${\mb S}^2$. 
\item[b)] Are globally defined on ${\mb H}^2$ for quadratic integrals (Koenigs case) while for higher even degree integrals they are {\bf never} globally defined.
\item[c)] Are globally defined on ${\mb H}^2$, under appropriate restrictions on the parameters, for integrals of odd power in the momenta.
\end{itemize}
\end{nTh}

\nin In all what follows the constraints $m_k>1$ will be assumed to hold.

%%%%%%%%%%%%%%%%%%%%%%%%%%%%%%%%%%%%%%%
\vspace{1cm}
\centerline{\large\bf I. INTEGRALS OF EVEN DEGREE}

%%%%%%%%%%%%%%%%%%%%%%%%%%%%%%%%%%%%%%%

\section{The setting}
Let us turn ourselves to the general case where 
$\sharp(S_1)=\sharp(S_2)=2n\geq 2$. The hamiltonian is
\beq\label{ham2}
H=\Pi^2+\frac{P_y^2}{\cosh^2 t}, \quad \Pi=\frac{P_t}{A(t)},\qq A(t)=1+\sum_{k=1}^{2n-1}\frac{e_k\,\sinh t}{\sqrt{m_k\,\cosh^2 t-1}},\eeq
and the extra integrals are given by
\beq
S_1=\cos y\,{\cal S}+\sin y\,{\cal T}, \qq S_2=\cos y\,{\cal T}-\sin y\,{\cal S},\eeq
where
\beq
{\cal S}=\sum_{k=0}^n\,\la_{2k-1}(t)\,H^{n-k}\,P_y^{2k},\qq\qq {\cal T}=\Pi\,\sum_{k=0}^{n-1}\,\la_{2k}(t)\,H^{n-k-1}\,P_y^{2k+1}.
\eeq
The main problem is therefore to determine the following array of functions of $t$ 
\beq
 \la_{-2}=0, \qq \la_{-1}=1 \qq  
\left(\barr{cccc}
\quad \la_1 & \quad \la_3 &\quad \ldots & \la_{2n-1}\\[4mm]
 \quad \la_0 & \quad \la_2 & \quad\ldots & \quad \la_{2(n-1)}\earr\right)\qq \la_{2n}=0.\eeq
The conventional values $\la_{-2}=\la_{2n}=0$ are introduced to alleviate many formulae in the sequel.

\section{Proof of Theorem 1}
Let us begin with
\begin{nth} $S_1$ and $S_2$ will be integrals iff the 
$\la$'s s solve the differential system:
\beq\label{sdiff1}
0\leq k \leq n: \qq \barr{clcl}
(a): \quad & \cosh^2 t\,\la'_{2k-1} & = & -A\,\la_{2(k-1)} \\[4mm]
(b): \quad & \cosh^2 t\,\la'_{2k} & = & \dst\la'_{2(k-1)}-\tanh t\,\la_{2(k-1)}-A\,\la_{2k-1},\earr
\eeq
taking into account the conventional values $\la_{-2}=\la_{2n}=0$.
\end{nth}

\vspace{5mm}
\nin{\bf Proof}: Both constraints $\{H,S_1\}=0$ and $\{H,S_2\}=0$ are seen to be equivalent to
\beq
\{H,{\cal S}\}=-2\frac{P_y}{\cosh^2 t}\,{\cal T} \qq\qq 
\{H,{\cal T}\}=-2\frac{P_y}{\cosh^2 t}\,{\cal S}.
\eeq
Using the explicit form of ${\cal S}$ and ${\cal T}$ elementary computations give (\ref{sdiff1}).$\hfill\Box$

A simplifying approach to the differential system (\ref{sdiff1}) makes use of generating functions, which encode all the array in a couple of objects:
\beq\label{defgf2}
{\cal L}(t,\xi)=\sum_{k=0}^{n-1}\la_{2k}(t)\,\xi^k \qq\qq {\cal M}(t,\xi)=\sum_{k=0}^n\la_{2k-1}(t)\,\xi^k.
\eeq

Let us prove
\begin{nth} The differential system given in Proposition 1 is equivalent, in terms of generating functions, to the set of partial differential equations
\beq
\cosh^2 t(1+\tau)\pt_t{\cal L} +\xi\,\tanh t\,{\cal L}+A\,{\cal M}=0, \qq \pt_t{\cal M}-\tau\,A\,{\cal L}=0,\eeq
where $\dst \tau=-\frac{\xi}{\cosh^2 t}.$
\end{nth}

\vspace{5mm}
\nin{\bf Proof:} Using relation (b) in (\ref{sdiff1}) we have
\beq
\cosh^2 t\,\pt_t{\cal L}=\sum_{k=0}^{n-1}\Big(\la'_{2(k-1)}-\tanh t\,\la_{2(k-1)}-A\,\la_{2k-1}\Big)\xi^k
\eeq
which becomes
\beq
\sum_{l=0}^{n-1}\Big(\la'_{2l}-\tanh t\,\la_{2l}\Big)\xi^{l+1}-\Big(\la'_{2(n-1)}-\tanh t\,\la_{2(n-1)}\Big)\xi^n-A\Big({\cal M}-\la_{2n-1}\,\xi^n\Big),\eeq
and can be written
\beq
\xi\pt_t{\cal L}-\xi\,\tanh t\,{\cal L}-A\,{\cal M}-\Big(\la'_{2(n-1)}-\tanh t\,\la_{2(n-1)}-A\,\la_{2n-1}\Big)\xi^n.
\eeq
So we have obtained
\beq
\cosh^2 t(1+\tau)\pt_t{\cal L} +\xi\,\tanh t\,{\cal L}+A\,{\cal M}=-\Big(\la'_{2(n-1)}-\tanh t\,\la_{2(n-1)}-A\,\la_{2n-1}\Big)\xi^n.
\eeq
The right hand member does vanish thanks to relation (b) for k = n in (\ref{sdiff1}). Conversely, expanding this relation in powers of $\xi$ one recovers relations (b) in (\ref{sdiff1}).

Using relation (a) in (\ref{sdiff1}) we have
\beq
\pt_t{\cal M}=\sum_{k=1}^n\,\la'_{2k-1}\,\xi^k
=-\frac{A}{\cosh^2 t}\,\sum_{l=0}^{n-1}\la_{2l}\,\xi^{l+1}=\tau\,A\,{\cal L},
\eeq
which was to be proved.$\hfill\Box$

To get an explicit form of these functions we need: 
\begin{ndef}\label{newH}
For $n\geq 1$ and $k\in\{1,2,\ldots,\nu\}$ let us introduce   
\beq\label{hk2}
\forall k: \quad h_k(t)=e_k\,\sqrt{m_k\,\cosh^2 t-1},\qq  e_k=\pm 1, \quad m_k>1,\eeq
and let us define the $H^{\nu}_k(t)$ by the 
\beq\label{gfH2}
{\cal H}^{\nu}(t,\xi)\equiv \prod_{k=1}^{\nu}(1+\xi\,h_k(t))=\sum_{k=0}^{\nu}\,H^{\nu}_k(t)\,\xi^k.
\eeq\end{ndef}

In Appendix A the reader will find useful relations for these functions to be used in the sequel. 

In this first part with $\sharp(S)=2n$ we will take $\nu=2n-1$ and to simplify  matters we will omit the upper index $\nu=2n-1$ of the functions $H^{2n-1}_k$.  

These functions allow to define the $\la$'s:
\begin{ndef}\label{sollambda2} Let us consider, for $k\in\{0,1,\ldots,n-1\}$ 
\beq\label{lapair2}
\la_{2k}=\frac{(-1)^{k+1}}{\cosh^{2k+1} t}\sum_{l=0}^k\,(-1)^l{n-1-l \choose n-1-k}\Big[H_{2l+1}+\sinh t\,H_{2l}\Big],
\eeq
for $k\in\,\{1,\ldots,n-1\}$ \footnote{This set is empty for $n=1$.}
\beq\label{laimpair21}
\la_{2k-1}=\frac{(-1)^k}{\cosh^{2k} t}\left\{\sum_{l=0}^k\,(-1)^l\,{n-l \choose n-k}\,H_{2l}-\sinh t\,\sum_{l=0}^{k-1}\,{n-1-l \choose n-k}\,H_{2l+1}
\right\},
\eeq
and for $k=n$ 
\beq\label{laimpair22}
\la_{2n-1}=\frac{(-1)^{n}}{\cosh^{2n} t}\sum_{l=0}^{n-1}\,(-1)^l\,\Big[H_{2l}-\sinh t\,H_{2l+1}\Big].\eeq
\end{ndef}
Let us compute the generating functions.

\begin{nth} 
Defining
\beq
\psi_{n,l}=\tau^l(1+\tau)^{n-l}, \qq  0\leq l \leq n,
\eeq
the generating functions are given by
\beq\label{gfL2}
{\cal L}(t,\xi)=-\frac 1{\cosh t}\,
\sum_{l=0}^{n-1}\,(-1)^l\,\psi_{n-1,l}\Big[H_{2l+1}+\sinh t\,H_{2l}\Big],
\eeq
and by
\beq\label{gfM2}
{\cal M}(t,\xi)=\sum_{l=0}^{n-1}\,(-1)^l\psi_{n,l}\,H_{2l}-\sinh t\,\sum_{l=0}^{n-1}(-1)^l\psi_{n,l+1}\,H_{2l+1}.
\eeq
\end{nth}

\vspace{5mm}
\nin{\bf Proof:} From the definition of ${\cal L}$, given in (\ref{defgf2}), and upon use of the formulae in (\ref{lapair2}) we have 
\beq
-\cosh t\,{\cal L}=\sum_{k=0}^{n-1}\tau^k\sum_{l=0}^k
(-1)^l{n-1-l \choose n-1-k}\Big[H_{2l+1}+\sinh t\,H_{2l}\Big].
\eeq
Reversing the order of the summations leads to
\beq
-\cosh t\,{\cal L}=\sum_{l=0}^{n-1}(-1)^l\Big[H_{2l+1}+\sinh t\,H_{2l}\Big]\sum_{k=l}^{n-1}{n-1-l \choose k-l}\tau^k,
\eeq
and since we have
\beq\label{binome}
\sum_{k=l}^{n-1}{n-1-l \choose k-l}\tau^k=\sum_{K=0}^{n-1-l}{n-1-l \choose K}\tau^{K+l}=\tau^l(1+\tau)^{n-1-l}=\psi_{n-1,l}\eeq
the relation (\ref{gfL2}) is proved.

From the definition of ${\cal M}$, given in  (\ref{defgf2}), and upon use of the formulae in (\ref{laimpair21}) and (\ref{laimpair22}) we have 
\beq
{\cal M}=\sum_{k=0}^n\tau^k\sum_{l=0}^k(-1)^l{n-l \choose n-k}H_{2l}-\sinh t\sum_{k=1}^n\tau^k\sum_{l=0}^{k-1}(-1)^l{n-1-l \choose n-k}H_{2l+1}.
\eeq
Reversing the summations gives
\beq
{\cal M}=\sum_{l=0}^n(-1)^l H_{2l}\sum_{k=l}^n{n-l \choose k-l}\tau^k-\sinh t\,\sum_{l=0}^{n-1}(-1)^l H_{2l+1}\sum_{k=l+1}^n {n-1-l \choose n}\tau^k.
\eeq
Using the binomial theorem, as explained in (\ref{binome}), gives
\beq\label{idbin}
\sum_{k=l}^n{n-l \choose k-l}\tau^k=\psi_{n,l}, \qq 
\sum_{k=l+1}^n {n-1-l \choose k-l-1}\tau^k=\psi_{n,l+1},
\eeq
which proves (\ref{gfM2}).$\hfill\Box$

Now let us use Proposition 3 to prove

\begin{nth}\label{ellemm} 
The generating functions ${\cal L}$ and ${\cal M}$ obtained in Proposition 3 are solutions of the partial differential equations given in Proposition 2, namely:
\beq\label{syscomplet}\left\{\barr{cl}
(a): & \qq \cosh^2 t(1+\tau)\pt_t{\cal L}+\xi\tanh t{\cal L}+A{\cal M}=0,\\[4mm]
(b): & \qq \pt{\cal M}-\tau\,A\,{\cal L}=0.\earr\right.\eeq
\end{nth} 

\vspace{5mm}
\nin{\bf Proof of relation (a):} Let us define the splitting
\beq
{\cal L}_1=-\frac 1{\cosh t}\sum_{l=0}^{n-1}\,(-1)^l\psi_{n-1,l}\, H_{2l+1}\qq\qq {\cal L}_2=-\tanh t\,\sum_{l=0}^{n-1}(-1)^l\psi_{n-1,l}\, H_{2l} \eeq
and similarly
\beq
{\cal M}_1=\sum_{l=0}^{n-1}(-1)^l\psi_{n,l}\, H_{2l}\qq\qq {\cal M}_2=-\sinh t\,\sum_{l=0}^{n-1}(-1)^l\psi_{n,l+1}\, H_{2l+1}.\eeq
In the sequel we will need the easily proved relations, valid for $l=0,1,\ldots,n-1$: 
\beq\label{easy}\left\{\barr{cl}
(a): & \quad \cosh t(1+\tau)\pt_t\psi_{n-1,l}=-\sinh t\Big[2l\,\psi_{n,l}+2(n-l-1)\,\psi_{n,l+1}\Big],\\[4mm]  (b): & \quad (1+\tau)\psi_{n-1,l}=\psi_{n,l},\\[4mm]
(c): & \quad \tau\,\psi_{n-1,l}=\psi_{n,l+1}. \earr\right.\eeq
Using relation (a) and the Proposition \ref{derH} in Appendix A, we get for $\ \cosh^2 t(1+\tau)\pt_t{\cal L}_1$:
\beq\barr{l}\dst 
\sinh t\sum_{l=0}^{n-1}(-1)^l\psi_{n,l} H_{2l+1}
+\sinh t\sum_{l=0}^{n-1}(-1)^l\Big[2l\,\psi_{n,l}+2(n-l-1)\,\psi_{n,l+1}\Big] H_{2l+1}\\[4mm]\dst 
-\sinh t\sum_{l=0}^{n-1}(-1)^l\Big[(2l+1)H_{2l+1}+(2l-2n)H_{2l-1}+\frac{(A-1)}{\sinh t}H_{2l}\Big]\psi_{n,l}.\earr
\eeq
All the terms involving $\psi_{n,l}\,H_{2l+1}$ add up to zero. The term involving $A-1$ is nothing but $-(A-1){\cal M}_1$ and the remaining terms
\beq
\sinh t\Big(\sum_{l=0}^{n-1}(-1)^l(2n-2l-2)\psi_{n,l+1}H_{2l+1}-\sum_{l=1}^n(-1)^l(2l-2n)\psi_{n,l}H_{2l-1}\Big)
\eeq
add up to zero. Hence we have proved
\beq
\cosh^2 t(1+\tau)\pt_t{\cal L}_1=-(A-1){\cal M}_1.
\eeq
The relation (c) in (\ref{easy}) implies that 
$\ \xi\,\tanh t\,{\cal L}_1=-{\cal M}_2$ so we conclude to
\beq\label{bout1}
\cosh^2 t(1+\tau)\pt_t{\cal L}_1+\xi\tanh t{\cal L}_1=-A{\cal M}_1+{\cal M}_1-{\cal M}_2.
\eeq

Similarly one can prove
\beq\label{bout2}
\cosh^2 t(1+\tau)\pt_t{\cal L}_2+\xi\tanh t{\cal L}_2=-A{\cal M}_2-{\cal M}_1+{\cal M}_2.
\eeq
Adding up (\ref{bout1}) and (\ref{bout2}) gives
\beq
\cosh^2 t(1+\tau)\pt_t{\cal L}+\xi\tanh t{\cal L}+A{\cal M}=0,\eeq
which is the relation (a) to be proved in Proposition \ref{ellemm}.

\vspace{5mm}
\nin {\bf Proof of relation (b):} Using the easy relation
\beq
\pt_t\psi_{n,l}=-\tanh t\Big[2l\,\psi_{n,l}+2(n-l)\psi_{n,l+1}\Big],\qq\quad 0\leq l \leq n,\eeq
and Proposition \ref{derH} we have
\beq\barr{l}\dst 
\frac 1{\tanh t}\pt_t{\cal M}_1=-\sum_{l=0}^{n-1}(-1)^l\Big[2l\,\psi_{n,l}+2(n-l)\psi_{n,l+1}\Big]H_{2l}+\sum_{l=0}^{n-1}(-1)^l\psi_{n,l}\,2l\,H_{2l}
\\[4mm]\dst 
+\sum_{l=1}^{n-1}(-1)^l\psi_{n,l}\Big[(2l-2n-1)H_{2(l-1)}+\frac{(A-1)}{\sinh t}H_{2l-1}\Big].\earr
\eeq
The terms involving $\psi_{n,l}\,H_{2l}$ add up to zero. In the second line the missing term for $l=n$  
\beq
-H_{2(n-1)}+\frac{(A-1)}{\sinh t}
\eeq
does vanish using (\ref{Hspe}), so let us add it. Observing that 
\beq
\tau{\cal L}_1=-\frac 1{\cosh t}\sum_{l=0}^{n-1}(-1)^l\tau\psi_{n-1,l}H_{2l+1}=-\frac 1{\cosh t}\sum_{l=0}^{n-1}(-1)^l\psi_{n,l+1}H_{2l+1},\eeq
the factor of $A-1$ becomes
\beq
\frac 1{\sinh t}\sum_{l=1}^n(-1)^l\psi_{n,l}H_{2l-1}=-\frac 1{\sinh t}\sum_{L=0}^{n-1}(-1)^l\psi_{n,L+1}H_{2L+1}=\frac 1{\tanh t}\,\tau\,{\cal L}_1.
\eeq
The remaining terms are
\beq\barr{l}\dst 
-\sum_{l=0}^{n-1}(-1)^l\,2(n-l)\psi_{n,l+1}H_{2l}+\sum_{l=1}^n(-1)^l\psi_{n,l}(2l-2n-1)H_{2(l-1)}=\\[4mm]
\dst =-\sum_{l=0}^{n-1}(-1)^l\psi_{n,l+1}H_{2l}=-\sum_{l=0}^{n-1}(-1)^l\,\tau\,\psi_{n-1,l}H_{2l}=\frac 1{\tanh t}\tau{\cal L}_2.\earr
\eeq
So we have obtained
\beq\label{M1}
\pt_t{\cal M}_1=A\,\tau\,{\cal L}_1-\tau\,{\cal L}_1+\tau\,{\cal L}_2.\eeq

Similarly one can prove
\beq\label{M2}
\pt_t{\cal M}_2=A\,\tau{\cal L}_2+\tau{\cal L}_1-\tau{\cal L}_2.\eeq
Adding (\ref{M1}) and (\ref{M2}) ends up the proof of relation (b) in Proposition \ref{ellemm}.
$\hfill\Box$

The generating functions ${\cal L}$ and 
${\cal M}$, given by (\ref{gfL2}) and (\ref{gfM2}), are solutions of the system of equations (\ref{syscomplet}). 
 It follows, from Proposition 2 that $S_1$ and $S_2$ are indeed integrals for the hamiltonian (\ref{ham2}). Its  geodesic flow is therefore superintegrable.
 
{\bf This concludes the proof of Theorem 1.}$\hfill\Box$ 

\section{Proof of Theorem 2}
\subsection{Computing the moments}
Let us define
\beq
S_+=S_1+S_2=e^y({\cal S}+{\cal T}),\qq\quad S_-=S_1-S_2=e^{-y}({\cal S}-{\cal T}).\eeq
It follows that we can define the moments $\si_k$\ according to
\beq\label{moms}
S_+\,S_-={\cal S}^2-{\cal T}^2=\sum_{k=0}^{2n}\si_k\,H^{2n-k}\,P_y^{2k}.\eeq
The moments are mere constants since $S_+\,S_-$ is an integral, and we have

\begin{nth} The moments are given in terms of the $\la$'s by the relations
\beq\label{formmom}
\barr{cll}
0\leq k \leq n: & \qq\dst \si_k=\sum_{l=0}^k\,S_{l,k-l} & \qq (\si_0=1)\\[4mm]
n+1 \leq k \leq 2n: & \qq\dst \si_k=\sum_{l=k-n}^n\,S_{l,k-l} & \earr
\eeq
where
\beq\label{Slk}
S_{l,k}=\la_{2l-1}\la_{2k-1}-\la_{2l}\la_{2(k-1)}+\frac{\la_{2(l-1)}\la_{2(k-1)}}{\cosh^2 t}.
\eeq
\end{nth}

\vspace{5mm}
\nin {\bf Proof:} Elementary computations involving products of finite series.$\hfill\Box$

In order to get more insight into the moments let us prove:

\begin{nth}\label{sigma} The generating function of the moments, defined by
\beq
\Si(\xi)=\sum_{k=0}^{2n}\si_k\,\xi^k
\eeq
is given in terms of the generating functions by
\beq\label{gfSi}
\Si={\cal M}^2-\xi(1+\tau){\cal L}^2.\eeq
\end{nth}

\vspace{5mm}
\nin {\bf Proof:} Using the relations given in  (\ref{formmom}) we have
\beq
\Si(\xi)=\sum_{l=0}^n\xi^l\sum_{k=0}^lS_{k,l-k}+\sum_{l=n+1}^{2n}\xi^l\sum_{k=l-n}S_{k,l-k}.
\eeq
Reversing the order of the summations gives
\beq
\Si=\sum_{k=0}^n\xi^k\sum_{l=k}^nS_{k,l-k}\xi^{l-k}+\sum_{k=1}^n\xi^k\sum_{l=n+1}^{k+n}S_{k,l-k}\xi^{l-k}=\sum_{l=0}^nS_{0,l}\xi^l+\sum_{k=1}^n\xi^k\sum_{l=k}^{k+n}S_{k,l-k}\xi^{l-k},
\eeq
so, defining $L=l-k$ we obtain
\beq
\Si=\sum_{k=0}^n\xi^k\sum_{L=0}^nS_{k,L}\xi^{L}=\sum_{k=0}^n\xi^k\sum_{L=0}^n\left(\la_{2k-1}\la_{2L-1}-\la_{2k}\la_{2(L-1)}+\frac{\la_{2(k-1)}\la_{2(L-1)}}{\cosh^2 t}\right)\xi^L
\eeq
Taking into account the conventional values: $\la_{-4}=\la_{-2}=\la_{2n}=0$, let us compute the first term 
\beq
\sum_{k=0}^n\la_{2k-1}\xi^k\,\sum_{L=0}^n\la_{2L-1}\xi^L={\cal M}^2.
\eeq
The second term is 
\beq
-\sum_{k=0}^{n-1}\la_{2k}\xi^k\sum_{L=1}^n\la_{2(L-1)}\xi^L=-{\cal L}\,\sum_{l=0}^{n-1}\la_{2l}\xi^{l+1}=-\xi{\cal L}^2,
\eeq
while the last term is
\beq
\frac 1{\cosh^2 t}\sum_{k=1}^n\la_{2(k-1)}\xi^k\sum_{L=1}^n\la_{2(L-1)}\xi^L=\frac 1{\cosh^2 t}\sum_{K=0}^{n-1}
\la_{2K}\xi^{K+1}\sum_{l=0}^n\la_{2l}\xi^{l+1}=-\xi\,\tau{\cal L}^2,
\eeq
which concludes the proof.$\hfill\Box$

We are now in position to establish

\begin{nth} We have the explicit relation
\beq\label{fctSi2}
\Si(\xi)=(1-\xi)\prod_{k=1}^{2n-1}(1-m_k\,\xi). 
\eeq
Using the symmetric functions $(M)_k$ of the masses 
$m_k$, defined by
\[ \prod_{k=1}^{2n-1}(1-m_k\,\xi)=\sum_{k=0}^{2n-1}(-1)^k(M)_k\,\xi^k, \]
one can express the moments according to
\beq\label{Simom}
\barr{l}
\si_0=1,\\[4mm]
\si_k=(-1)^k\Big[(M)_k+(M)_{k-1}\Big], \qq 1 \leq k \leq 2n-1, \\[4mm]\dst \si_{2n}=(M)_{2n-1}=\prod_{k=1}^{2n-1} m_k.\earr\eeq
\end{nth}

\vspace{5mm}
\nin {\bf Proof:} The second part of this Proposition is trivial:  expanding $\Si(\xi)$ in powers of $\xi$ gives (\ref{Simom}).

In order to prove (\ref{fctSi2}) we need a new writing of the generating functions. Let us begin with
\beq
{\cal M}=\sum_{l=0}^{n-1}(-1)^l\psi_{n,l}H_{2l}-\sinh t
\sum_{l=0}^{n-1}(-1)^l\psi_{n,l+1}H_{2l+1}.
\eeq
For $\xi<0$ one can define $\dst \eta=\sqrt{\frac{\tau}{1+\tau}}$ which gives
\beq
{\cal M}=(1+\tau)^n\left(\sum_{l=0}^{n-1}(-1)^l\,\eta^{2l}\,H_{2l}-\eta\sinh t\,\sum_{l=0}^{n-1}(-1)^l\,\eta^{2l+1}\,H_{2l+1}\right).
\eeq
Defining
\beq
{\cal H}_{\pm}={\cal H}(t,\pm i\eta)=\prod_{k=1}^{2n-1}\Big(1 \pm i\eta h_k(t)\Big),\qq h_k=e_k\sqrt{m_k\,\cosh^2 t-1}
\eeq
and using the relations given in (\ref{Hpm}) of Appendix A, we get
\beq
{\cal M}=\frac{(1+\tau)^n}{2}\Big((1+i \eta \sinh t){\cal H}_++(1-i\eta \sinh t) {\cal H}_-\Big).
\eeq
Similarly we have
\beq
{\cal L}=i\frac{(1+\tau)^{n-1}}{2\eta\cosh t}\Big((1+i\eta\sinh t){\cal H}_+-(1-i\eta \sinh t){\cal H}_-\Big).
\eeq
Plugging these relations into (\ref{gfSi}) we deduce
\beq
\Si(\xi)=(1+\eta^2\sinh^2 t)\,(1+\tau)^{2n}{\cal H}_+{\cal H}_-,\eeq
and the relations 
\beq
1+\eta^2\sinh^2 t=\frac{1-\xi}{1+\tau},\qq (1+\tau)^{2n}{\cal H}_+{\cal H}_-=(1+\tau)\prod_{k=1}^{2n-1}(1-\xi\,m_k),
\eeq
prove (\ref{fctSi2}) for $\xi<0$. Since $\Si$ is a polynomial the result is valid for any complex $\xi$.
$\hfill\Box$

\subsection{Poisson algebra}
To establish the Poisson algebra structure we need to compute
\beq
-\frac 12\{S_+,S_-\}=\frac 12\frac{\pt}{\pt P_y}({\cal S}^2-{\cal T}^2)+\{{\cal S},{\cal T}\}.
\eeq
The first term is
\beq
\sum_{k=1}^{2n}k\si_k\,H^{2n-k}\,P_y^{2k-1}+\frac 1{\cosh^2 t}\sum_{k=0}^{2n}(2n-k)\si_k\,H^{2n-l-1}
\,P_y^{2l+1},
\eeq
so if we prove the relation
\beq\label{miracle}
\cosh^2 t\,\{{\cal S},{\cal T}\}+\sum_{k=0}^{2n}(2n-k)\si_k\,H^{2n-k-1}\,P_y^{2k+1}=0,
\eeq
we will have proved Theorem 2
\beq
\{S_+,S_-\}=-2\sum_{k=0}^{2n-1}(k+1)\si_{k+1}\,H^{2n-k-1}\,P_y^{2k+1}.\eeq

Let us proceed to
\begin{nth} The relation (\ref{miracle}) does hold true.
\end{nth}

\vspace{3mm}
\nin{\bf Proof:} Let us compute the Poisson bracket
\beq
\{{\cal S},{\cal T}\}=\sum_{k,l}P_y^{2(k+l)-1}\{\la_{2k-1}\,H^{n-k},\la_{2(l-1)}\,\Pi\,H^{n-l}\}.
\eeq
Introducing the notation $\dst \Psi^{2n}_{k+l}=H^{2n-k-l-1}\,P_y^{2(k+l)-1}$ we get for the right hand side \footnote{Any mute index is summed from $0$ to $n$.}
\beq
\barr{l}\dst 
\sum_{k,l}\Psi^{2n}_{k+l}\Big((n-k)\la_{2k-1}\{H,\la_{2(l-1)}\}\Pi+(n-k)\la_{2k-1}\la_{2(l-1)}\{H,\Pi\}-\\[4mm]
\hspace{4cm}-(n-k)\la_{2(k-1)}\{H,\la_{2l-1}\}\Pi-\la_{2(l-1)}\{\Pi,\la_{2k-1}\}H\Big).\earr
\eeq
The remaining brackets are all elementary and lead to
\beq
\barr{l}\dst 
\sum_{k,l}\Psi^{2n}_{k+l}\Big(2(n-k)\la_{2k-1}\frac{\la'_{2(l-1)}}{A}\,\Pi^2+2(n-k)\frac{\la_{2(k-1)}\la_{2(l-1)}}{\cosh^2 t}\,\Pi^2+\\[4mm]\dst 
\hspace{4cm}+2(n-k)\tanh t\,\frac{\la_{2k-1}\la_{2(l-1)}}{\cosh^2 t\,A}P_y^2+\frac{\la_{2(k-1)}\la_{2(l-1)}}{\cosh^2 t}H\Big).
\earr\eeq
Getting rid of $\dst \Pi^2=H-\frac{P_y^2}{\cosh^2 t}$ and upon multiplication by $\cosh^2 t$ one obtains
\beq\label{interm}
\barr{l}\dst 
\sum_{k,l}\Psi^{2n}_{k+l+1}\left[2(n-k)\la_{2k-1}\frac{\cosh^2 t\la'_{2(l-1)}}{A}+(2n-2k+1)\la_{2(k-1)}\la_{2(l-1)}\right]+\\[6mm]\dst 
+\sum_{k,l}\Psi^{2n}_{k+l}\left[2(n-k)\frac{\la_{2k-1}}{A}\Big(-\la'_{2(l-1)}+\tanh t\,\la_{2(l-1)}\Big)-2(n-k)\frac{\la_{2(k-1)}\la_{2(l-1)}}{\cosh^2 t}\right].\earr
\eeq
Let us consider the first term
\beq
\sum_{k}\,2(n-k)\la_{2k-1}\sum_{l=1}^n\Psi^{2n}_{k+l+1}\frac{\cosh^2 t\la'_{2(l-1)}}{A}=\sum_{k}\,2(n-k)\la_{2k-1}\sum_{L=0}^n\Psi^{2n}_{k+L}\frac{\cosh^2 t\la'_{2L}}{A}.\eeq
Adding it to the similar terms in the second line of (\ref{interm}) we get
\beq
\barr{l}\dst 
\sum_{k,l}\Psi^{2n}_{k+l}\,2(n-k)\frac{\la_{2k-1}}{A}\Big(\cosh^2 t\,\la'_{2l}-\la'_{2(l-1)}+\tanh t\,\la_{2(l-1)}\Big)=\\[4mm]\dst 
\hspace{4cm}=-\sum_{k,l}\Psi^{2n}_{k+l}\,2(n-k)\la_{2k-1}\la_{2l-1}\earr
\eeq
upon use of relation (b) in (\ref{sdiff1}). Taking into account the $k\leftrightarrow l$ symmetry we can write the remaining terms
\beq
\barr{l}\dst 
\sum_{k,l}\Psi^{2n}_{k+l+1}\,(2n-k-l+1)\la_{2(k-1)}\la_{2(l-1)}-\\[4mm]\dst \hspace{4cm}-\sum_{k,l}\Psi^{2n}_{k+l}\,(2n-k-l)\left(\la_{2k-1}\la_{2l-1}+\frac{\la_{2(k-1)}\la_{2(l-1)}}{\cosh^2 t}\right).\earr
\eeq
The first term becomes
\beq
\sum_{K,l}\Psi^{2n}_{K+l}\,(2n-K-l)\la_{2K}\la_{2(l-1)},
\eeq
recalling the definition of $S_{k,l}$, given by (\ref{Slk}), we conclude to
\beq
\cosh^2 t\,\{{\cal S},{\cal T}\}=-\sum_{k,l}(2n-k-l)S_{k,l}\,H^{2n-k-l-1}P_y^{2(k+l)+1}.
\eeq
Setting $L=l+k$ we have
\beq
\cosh^2 t\,\{{\cal S},{\cal T}\}=-\sum_{k=0}^n\sum_{L=k}^{n+k}(2n-L)S_{k,L-k}\,H^{2n-L-1}\,P_y^{2L+1},
\eeq
and reversing the summations gives
\beq
\barr{l}\dst 
\cosh^2 t\,\{{\cal S},{\cal T}\}=-\sum_{L=0}^n\,(2n-L)\,H^{2n-L-1}\,P_y^{2L+1}\sum_{k=0}^L\,S_{k,L-k}-\\[4mm]\dst    \hspace{4cm}-\sum_{L=n+1}^{2n}(2n-L)\,H^{2n-L-1}\,P_y^{2L+1}\sum_{k=L-n}^nS_{k,L-k}.\earr
\eeq
Thanks to relations (\ref{formmom}) we have
\beq
\cosh^2 t\,\{{\cal S},{\cal T}\}=-\sum_{l=0}^{2n}(2n-l)\si_l\,H^{2n-l-1}\,P_y^{2l+1},
\eeq
which was to be proved. $\hfill\Box$

\vspace{2mm}
\nin{\bf This concludes the proof of Theorem 2.}$\hfill\Box$

\section{Relation with Koenigs}
For $n=1$ the integrals are quadratic in the momenta. This case was first solved by Koenigs in \cite{Ko}: 
\beq\left\{\barr{l} \dst 
H_K=\frac{\sin^2 x}{1-\rho\,\cos x}(P_x^2+P_y^2),\\[4mm]
S_1^K=\cosh y\Big(\frac{\rho}{2}\,H-\cos x\,P_y^2\Big)-\sinh y\,\sin x\,P_x\,P_y.\earr\right.
\eeq
As shown in \cite{Va1}, taking $x\in(0,\pi)$ and $y\in{\mb R}$ and defining $e^{\chi}=\tan(x/2)$, we obtain
\beq
\left\{\barr{l}\dst  
H_K=\frac1{1+\rho\,\tanh\chi}\left(P_{\chi}^2+\frac{P_y^2}{\cosh^2\chi}\right),\qq (\chi,\,y)\in{\mb R}^2,\quad \rho\in(0,+1),\\[4mm] 
S_1^K=\cosh y\Big(\frac{\rho}{2}\,H+\tanh\chi\,P_y^2\Big)-\sinh y\,P_{\chi}\,P_y.\earr\right.
\eeq 

Let us relate these results with our work. In Koenigs case the extra integral is quadratic so $n=1$ and    Theorem 1 gives for hamiltonian    
\beq
H=\Pi^2+\frac{P_y^2}{\cosh^2 t},\quad \Pi=\frac{P_t}{A},
\eeq
where
\beq
A=1+\frac{\sinh t}{\sqrt{m\cosh^2 t-1}},\qq (t,y)\in{\mb R}^2, \qq m>1.
\eeq
The first extra integral is
\beq
S_1=\cosh y\Big(H+\la_1\,P_y^2\Big)+\sinh y\,\la_0\,\Pi\,P_y,\eeq
and Definition \ref{sollambda2} gives
\beq
\la_0=-\frac 1{\cosh t}\Big(H_1+\sinh t\Big),\qq\quad \la_1=-\frac 1{\cosh t}\Big(1-\sinh t\,H_1\Big).
\eeq
Here we have $H_1=\sqrt{m\cosh^2 t-1}$ so that
\beq\label{intK}
\barr{l} \dst 
S_1=\cosh y\left(H+\frac{(\sinh t\,\sqrt{m\cosh^2 t-1}-1)}{\cosh^2 t}\,P_y^2\right)-\\[4mm]\dst 
\hspace{5cm} -\sinh y\,\frac{(\sinh t+\sqrt{m\cosh^2 t-1})}{\cosh t}\,\Pi\,P_y.\earr\eeq
Let us describe the diffeomorphism relating $H_K$ and $H$:

\begin{nth} The correspondence
\beq\label{echi}
e^{\chi}=\frac{\sqrt{m}\sinh t+\sqrt{m\cosh^2 t-1}}{\sqrt{m}+1},\qq\quad \chi\in{\mb R},\eeq
implies the relations
\beq
H_K=\frac H{\mu^2}, \qq\qq S_1^K=S_1,\qq\qq \mu=\sqrt{\frac{m}{m+1}}.\eeq
\end{nth}

\vspace{5mm}\nin{\bf Proof:} The relation $g_K=\mu^2\,g$ is equivalent to
\beq
(a): \quad\sqrt{1+\rho\tanh\chi}\,\cosh\chi=\mu\cosh t \quad \&\quad  (b): \quad \sqrt{1+\rho\tanh\chi}\,d\chi=\mu\,A\,dt.\eeq
Taking $\dst \rho=\frac{2\sqrt{m}}{m+1} $ the relation (a) is purely algebraic and has for solution
\[e^{2\chi}=\frac{2m\cosh^2 t-m-1+2\sqrt{m}\sinh t\,a(t)}{(\sqrt{m}+1)^2}.\]
Taking its square root gives (\ref{echi}) and 
$\dst \frac{d\chi}{dt}=\frac{\sqrt{m}\cosh t}{\sqrt{m\cosh^2 t-1}}.$ 
A somewhat hairy computation gives then
\[\sqrt{m+1}\sqrt{1+\rho\tanh\chi}=\frac{\sinh t+\sqrt{m\cosh^2 t-1}}{\cosh t},\]
which shows that the relation (b) is identically true.
$\hfill\Box$

%%%%%%%%%%%%%%%%%%%%%%%%%%%%%%%%%%%%%%%
\newpage
\centerline{\large\bf II. INTEGRALS OF ODD DEGREE}

%%%%%%%%%%%%%%%%%%%%%%%%%%%%%%%%%%%%%%%

\vspace{3mm}
\section{The setting}
Here $\sharp(S_1)=\sharp(S_2)=2n+1\geq 3$. 
Let us recall that the hamiltonian is 
\beq\label{hamimpair}
H=\Pi^2+\frac{P_y^2}{\cosh^2 t},\quad  
\Pi=\frac{P_t}{A(t)}, \qq A(t)=1+\sum_{k=1}^{2n}\frac{e_k\,\sinh t}{\sqrt{m_k\,\cosh^2 t-1}}\eeq
and the extra integrals 
\beq\label{intodd}
S_1=\cosh y\,{\cal S}+\sinh y\,{\cal T},\qq S_2=\{P_y,S_1\}=\cosh y\,{\cal T}+\sinh y\,{\cal S}.
\eeq
The following array of functions of $t$:
\[\la_{-1}=1,\quad\left(\barr{cccc}
\la_1 &\quad \la_3 &\quad \ldots & \la_{2n}\\[4mm]
\la_0 & \quad \la_2  & \quad\ldots & \quad \la_{2n-1}\earr\right), \quad \la_{2n+1}=0,\]
allows to define the building blocks of $S_1$ and $S_2$:
\beq\label{AB1}
{\cal S}=\Pi\sum_{k=0}^n\,\la_{2k-1}(t)\,H^{n-k}\,P_y^{2k}, \qq\qq {\cal T}=\sum_{k=0}^n\,\la_{2k}(t)\,H^{n-k}\,P_y^{2k+1}.
\eeq  

\section{Proof of Theorem 3}

Let us begin with:
\begin{nth}
$S_1$ and $S_2$ will be integrals if and only if the 
$\la$'s solve the differential system:
\beq\label{sdiff2}0\leq k \leq n: \qq 
\barr{cccl}
(a): & \qq \cosh^2 t\,\la'_{2k} & = & -A\,\la_{2k-1},
\\[4mm]
(b): & \qq \cosh^2 t\,\la'_{2k+1} & = & \dst \la'_{2k-1}-\tanh t\,\la_{2k-1}-A\,\la_{2k}. 
\earr\eeq
\end{nth}

\vspace{3mm}
\nin{\bf Proof:} Similar to the proof of Proposition 1.
$\hfill\Box$

Let us define the generating functions defined by
\beq
{\cal L}(t,\xi)=\sum_{k=0}^n\,\la_{2k}(t)\,\xi^k,\qq\quad {\cal M}(t,\xi)=\sum_{k=0}^n\,\la_{2k-1}(t)\,\xi^k,\qq\xi\in{\mb C},
\eeq
and let us prove
\begin{nth} The differential system in Proposition 10 is equivalent to the partial differential system for the generating functions:
\beq
\cosh^2 t\,\pt_t{\cal L}+A{\cal M}=0,\qq 
\cosh^2 t(1+\tau)\,\pt_t{\cal M}+\xi\tanh t\,{\cal M}+\xi\,A\,{\cal L}=0,\eeq
where $\dst \tau=-\frac{\xi}{\cosh^2 t}.$
\end{nth}

\vspace{3mm}
\nin{\bf Proof:} Using relation (a) in (\ref{sdiff2}) we have
\beq
\cosh^2 t\,\pt_t{\cal L}=-A\sum_{k=0}^n\la_{2k-1}\xi^k=-A\,{\cal M}.
\eeq
Conversely, expanding this relation in powers of $\xi$ gives relation (a) in (\ref{sdiff2}).

Using relation (b) in (\ref{sdiff2}) we have
\beq
\cosh^2 t\,\pt_t{\cal M}=\sum_{k=0}^{n-1}\xi^{k+1}\Big(\la'_{2k-1}-\tanh t\,\la_{2k-1}-A\,\la_{2k}\Big),
\eeq
or in terms of the generating functions
\beq
\cosh^2 t\,\pt_t{\cal M}=\xi\Big(\pt_t\,{\cal M}-\tanh t\,{\cal M}-A\,{\cal L}\Big)-\xi^{n+1}(\la'_{2n-1}-\tanh t\,\la_{2n-1}-A\,\la_{2n}),
\eeq
leading to
\beq
\cosh^2 t\,(1+\tau)\pt_t{\cal M}+\xi\tanh t{\cal M}+\xi A {\cal L}=-\xi^{n+1}(\la'_{2n-1}-\tanh t\,\la_{2n-1}-A\,\la_{2n}).\eeq
The right hand side of this relation does vanish due to 
relation (b) for $k=n$. Conversely, expanding this relation in powers of $\xi$ one recovers relation (b).
$\hfill\Box$

We will need again the functions of Definition \ref{newH}, but in the case $\sharp(S)=2n$ considered in this second part, we will take $\nu=2n-1$ and to simplify  matters we will omit the upper index $\nu=2n-1$ of the functions $H^{2n-1}_k$ from the formulae. 

These functions allow to define the $\la$'s:

\begin{ndef}
For $k\in\{0,1,\ldots,n-1\},$ let us take for the functions defining ${\cal S}$:
\beq
\quad \la_{2k}=\frac{(-1)^{k+1}}{\cosh^{2k+1} t}\sum_{l=0}^k\,(-1)^l\,{n-l \choose n-k}\,\Big[H_{2l+1}+\sinh t\,H_{2l}\Big],
\eeq
and
\beq
\la_{2n}=\frac{(-1)^{n+1}}{\cosh^{2n+1} t}\left[\sum_{l=0}^{n-1}\,(-1)^l H_{2l+1}+\sinh t\,\sum_{l=0}^n(-1)^l H_{2l}\right].
\eeq
The functions needed for ${\cal T}$ are, for $k\in\{1,2,\ldots,n-1\},$ given by 
\beq
\la_{2k-1}=\frac{(-1)^k}{\cosh^{2k} t}\left[\sum_{l=0}^k\,(-1)^l{n-l \choose n-k}\,H_{2l}-\sinh t\,\sum_{l=1}^{k-1}(-1)^l\,{n-1-l \choose n-k}\,H_{2l+1}\right],
\eeq
and
\beq
\la_{2n-1}=\frac{(-1)^n}{\cosh^{2n} t}\left[\sum_{l=0}^{n-1}\,(-1)^l\,H_{2l}-\sinh t\,\sum_{l=1}^{n-1}(-1)^l\,H_{2l+1}\right].
\eeq
\end{ndef}
Now it is possible to compute the generating functions 
${\cal L}$ and ${\cal M}$ :

\begin{nth} Defining
\beq
\psi_{n,l}=\tau^l(1+\tau)^{n-l}, \qq\quad 0\leq l \leq n,
\eeq
the generating functions are given by
\beq
-\cosh t\,{\cal L}(t,\xi)=\sum_{l=0}^{n-1}\,(-1)^l\psi_{n,l}H_{2l+1}+\sinh t\,\sum_{l=0}^n(-1)^l\psi_{n,l}\,H_{2l}
\eeq
and by
\beq
{\cal M}(t,\xi)=\sum_{l=0}^n\,(-1)^l\,\psi_{n,l}H_{2l}-\sinh t\,\sum_{l=0}^{n-1}(-1)^l\psi_{n,l+1}\,H_{2l+1}.
\eeq\end{nth}

\vspace{3mm}
\nin{\bf Proof:} We have for the first generating function
\beq\barr{l}\dst 
-\cosh t\,{\cal L}=\sum_{k=0}^{n-1}\tau^k\sum_{l=0}^k (-1)^l{n-l \choose n-k}[H_{2l+1}+\sinh t\,H_{2l}]+
\\[4mm]\dst 
\hspace{4cm}+\tau^n\sum_{l=0}^{n-1}(-1)^l[H_{2l+1}+\sinh t\,H_{2l}].\earr
\eeq
Reversing the summations in the first term we get
\beq
-\cosh t\,{\cal L}=\sum_{l=0}^{n-1}(-1)^l[H_{2l+1}+\sinh t\,H_{2l}]\sum_{k=l}^n{ n-l \choose n-k}\tau^k,
\eeq
and the first relation in (\ref{idbin}) concludes the proof.

For the second generating function we have
\beq
{\cal M}=\sum_{k=0}^n\tau^k\sum_{l=0}^k(-1)^l{n-l \choose n-k}H_{2l}-\sinh t\sum_{k=1}^n\tau^k\sum_{l=0}^{k-1}(-1)^l{n-1-l \choose n-k}H_{2l+1},
\eeq
and the first term, thanks to the first relation in (\ref{idbin}), gives
\beq
\sum_{l=0}^n(-1)^l H_{2l}\sum_{k=l}^n{n-l \choose n-k}\tau^k=\sum_{l=0}^n(-1)^l\,\psi_{n,l}\,H_{2l}.
\eeq
The second piece gives, thanks to the second relation in (\ref{idbin}):
\beq
-\sinh t\sum_{l=0}^{n-1}(-1)^l H_{2l+1}\sum_{k=l+1}^n{ n-1-l \choose k-l-1}\tau^k=-\sinh t\sum_{l=0}^{n-1}(-1)^l\psi_{n,l+1}\,H_{2l+1},
\eeq
which was to be proved.$\hfill\Box$

Let us proceed to
\begin{nth} The generating functions obtained in Proposition 12 are solutions of the partial differential equations given in Proposition 11, namely
\beq\label{sdiff2}\barr{cl}\dst 
(a): & \qq \cosh^2 t\,\pt_t{\cal L}+A{\cal M}=0,\\[4mm]\dst
(b): & \qq \cosh^2 t(1+\tau)\,\pt_t{\cal M}+\xi\tanh t\,{\cal M}+\xi\,A\,{\cal L}=0,\earr\eeq
\end{nth}

\vspace{3mm}
\nin{\bf Proof of relation (a):} Let us define the splittings
\beq
{\cal L}_1=-\frac 1{\cosh t}\sum_{l=0}^{n-1}(-1)^l\psi_{n,l}\,H_{2l+1}, \qq {\cal L}_2=-\tanh t\sum_{l=0}
^{n-1}(-1)^l\psi_{n,l}H_{2l},
\eeq
and
\beq
{\cal M}_1=\sum_{l=0}^{n}(-1)^l\psi_{n,l}\,H_{2l}, \qq {\cal M}_2=-\sinh t\sum_{l=0}^{n-1}(-1)^l\psi_{n,l+1}H_{2l+1}.
\eeq
Upon use of the easy relation
\[ \cosh^2 t\,\pt_t\psi_{n,l}=-\sinh t\cosh t(2l\,\psi_{n,l}+2(n-l)\psi_{n,l+1},\]
and of relation (\ref{derH}) in Appendix A, one gets
\beq\barr{l}\dst 
\frac{\cosh^2 t}{\sinh t}\,\pt_t\,{\cal L}_1=\sum_{l=0}^{n-1}(-1)^l\psi_{n,l}H_{2l+1}+\sum_{l=0}^{n-1}(2l\psi_{n,l}+2(n-l)\psi_{n,l+1})H_{2l+1}-\\[4mm]\dst 
\hspace{1cm}-\sum_{l=0}^{n-1}(-1)^l\psi_{n,l}((2l+1)H_{2l+1}-(2n-2l+1)H_{2l-1}+\frac{(A-1)}{\sinh t}(H)_{2l}.\earr
\eeq
The terms involving $\psi_{n,l}H_{2l+1}$ add up to zero. The term involving $(A-1)$ is
\beq
-\frac{(A-1)}{\sinh t}\sum_{l=0}^{n-1}(-1)^l\psi_{n,l}H_{2l}
=-\frac{(A-1)}{\sinh t}\Big({\cal M}_1-(-1)^n H_{2n}\Big).
\eeq
The remaining terms compensate partially and we are left with
\beq
\sum_{l=0}^{n-2}(-1)^l\psi_{n,l+1}H_{2l+1}=\frac 1{\sinh t}\Big(-{\cal M}_2+(-1)^{n-1}\sinh t H_{2n-1}\Big).
\eeq
Collecting all the terms we have obtained
\beq
\cosh^2 t\,\pt_t\,{\cal L}_1=-A{\cal M}_1+{\cal M}_1-{\cal M}_2+(-1)^{n-1}\Big(\sinh t\,H_{2n-1}-(A-1)H_{2n}\Big),
\eeq
adn the last term vanishes thanks to relation (\ref{Hspe}).

Similarly one can check that
\beq
\cosh^2 t\,\pt_t\,{\cal L}_2=-A{\cal M}_2-{\cal M}_1+{\cal M}_2.\eeq
Adding these last two equations proves relation $(a)$.

\vspace{3mm}
\nin{\bf Proof of relation (b):} Using the easy relations
\beq
\barr{l}
\cosh^2 t(1+\tau)\pt_t\psi_{n,l}=\xi\tanh t(2l\psi_{n,l-1}+2(n-l)\psi_{n,l}),\\[4mm] 
\cosh^2 t(1+\tau)\psi_{n,l-1}=-\xi\,\psi_{n,l-1},\earr
\eeq
one gets
\beq\barr{l}\dst 
\frac{\cosh^2 t}{\tanh t}(1+\tau)\,\pt_t\,{\cal M}_1=\xi\sum_{l=0}^n(-1)^l\Big(2l\psi_{n,l-1}+2(n-l)\psi_{n,l}\Big)H_{2l}-\\[4mm]\dst 
\hspace{4cm}-\xi\sum_{l=1}^n(-1)^l\psi_{n,l-1}\Big(2l  H_{2l}+(2l-2n-2)H_{2(l-1)}\Big)-\\[4mm]\dst 
\hspace{4cm}-\xi\frac{(A-1)}{\sinh t}\sum_{l=1}^n(-1)^l\psi_{n,l-1}H_{2l-1}.\earr
\eeq
The terms involving $\psi_{n,l-1}H_{2l}$ add up to zero. The term involving $(A-1)$ is
\beq
\xi\frac{(A-1)}{\sinh t}\sum_{l=0}^{n-1}(-1)^l\psi_{n,l}H_{2l+1}=-\frac{(A-1)}{\tanh t}{\cal L}_1.\eeq
The remaining terms add up to zero, so we have obtained
\beq
\cosh^2 t(1+\tau)\,\pt_t\,{\cal M}_1=-\xi\, A\,{\cal L}_1+\xi\,{\cal L}_1.
\eeq
Adding the relation $\ \xi\,\tanh t\,{\cal M}_1=-\xi{
\cal L}_2\ $ we end up with
\beq
\cosh^2 t(1+\tau)\,\pt_t\,{\cal M}_1+\xi\,\tanh t\,{\cal M}_1=-\xi\, A\,{\cal L}_1+\xi({\cal L}_1-{\cal L}_2).
\eeq
Similarly one can show the relation
\beq
\cosh^2 t(1+\tau)\,\pt_t\,{\cal M}_2+\xi\,\tanh t\,{\cal M}_2=-\xi\, A\,{\cal L}_2-\xi({\cal L}_1-{\cal L}_2).
\eeq
Adding these last two equations proves relation $(b)$.
$\hfill\Box$

So we have proved that the generating functions ${\cal L}$ and ${\cal M}$, as defined in Proposition 12 are indeed solutions of the partial differential relations stated in Proposition 11 which insure that $S_1$ and $S_2$ are integrals for the hamiltonian (\ref{hamimpair}). Its  geodesic flow is therefore superintegrable.
 
{\bf This concludes the proof of Theorem 3.}$\hfill\Box$

\section{The moments and the Poisson structure}
To avoid agony for the reader, we will state the results without proofs since the techniques and the relations established in the first part are easily adapted to this case.

Let us define
\beq
S_+=S_1+S_2=e^y({\cal S}+{\cal T}),\qq\quad S_-=S_1-S_2=e^{-y}({\cal S}-{\cal T}),\eeq
we have
\beq
S_+\,S_-={\cal S}^2-{\cal T}^2=\sum_{k=0}^{2n+1}\si_k\,H^{2n+1-k}\,P_y^{2k}.\eeq

The moments are given in terms of the $\la$'s by the relations
\beq
\barr{cll}
0\leq k \leq n: & \qq\dst \si_k=\sum_{l=0}^k\,S_{l,k-l} & \qq (\si_0=1)\\[4mm]
n+1 \leq k \leq 2n+1: & \qq\dst \si_k=\sum_{l=k-n-1}^n\,S_{l,k-l} & \earr
\eeq
where
\beq
S_{l,k}=\la_{2l-1}\la_{2k-1}-\la_{2l}\la_{2(k-1)}-\frac{\la_{2l-1}\la_{2k-3}}{\cosh^2 t}.
\eeq

Defining the generating function of the moments by
\beq
\Si(\xi)=\sum_{k=0}^{2n+1}\si_k\,\xi^k,\eeq
it is related to the generating functions by
\beq
\Si=(1+\tau){\cal M}^2-\xi\,{\cal L}^2.\eeq
Its computation gives
\beq
\Si(\xi)=(1-\xi)\prod_{k=1}^{2n}(1-\xi\,m_k).\eeq

The Poisson algebra structure follows from
\beq
\barr{l}\dst 
S_+\,S_-=\sum_{k=0}^{2n+1}\si_k\,H^{2n+1-k}\,P_y^{2k},\\[4mm]\dst 
\{S_+,S_-\}=-2\sum_{k=0}^{2n}(k+1)\si_{k+1}\,H^{2n-k}\,P_y^{2k+1},\earr
\eeq
where the moments are given by
\beq
\left\{\barr{lcll}
\si_0 & = & 1 & \\[4mm]
\si_l & = & (-1)^l\Big((M)_l+(M)_{l-1}\Big) & \quad l\in\{1,\ldots,2n \}\\[4mm]
\si_{2n+1} & = & \dst -(M)_{2n}=-\prod_{k=1}^{2n}m_k, & \earr\right.
 \eeq
where the symmetric functions $(M)_k$ of the $m_k$ are defined by
\beq
\prod_{k=1}^{2n}(1-\xi\,m_k)=\sum_{k=0}^{2n}\,(-1)^k(M)_k\,\xi^k.
\eeq
This concludes the statements of Theorem 4.$\hfill\Box$

Let us turn ourselves to the global problems.

\section{Global aspects}
The metric to discuss for $\nu\in{\mb N}\bs \{0\}$ is  
\beq\label{globmet}
g=A^2(t)\,dt^2+\cosh^2 t\,dy^2,\eeq
where 
\beq 
A(t)=1+\sum_{k=1}^{\nu}\frac{e_k\,\sinh t}{\sqrt{m_k\,\cosh^2 t-1}},\qq \forall k: \ e_k=\pm 1,\quad  m_k>1.\eeq
For integrals with $\sharp(S_i)=2n$ (resp. 
$\sharp(S_i)=2n+1$) one has to take $\nu=2n-1$ (resp. $\nu=2n$). For $\nu=1$ we recover Koenigs. 

\subsection{Proof of point a) in Theorem 5}
It is necessary that the sectional curvature be at least continuous for $t\in{\mb R}$. 
Since we have
\beq
R=\frac{\sinh t\,A'-\cosh t\,A}{A^3},\eeq
it follows that $A$ must not vanish, so we can take $A>0$ for $t\in{\mb R}$.

Let us compute the area of the surface:
\beq
\mu(M)=\int_{\mb R}\cosh t\,A(t)\,dt\,\int dy.\eeq
If one takes $y\in{\mb R}$ the second integral diverges. But this is not mandatory since we can take $y\in{\mb S}^1$ and the second integral gives $2\pi$. In this case let us consider the integral over $t$. Its convergence requires  
\[A(-\nf)=A(+\nf)=0\qq\Longleftrightarrow\qq  1\pm\sum_{k=1}^n\frac{e_k}{\sqrt{m_k}}=0\]
which is impossible, hence $\mu(M)$ is divergent excluding $M={\mb S}^2$.  $\hfill\Box$

\subsection{Conformal structure of the metric}
To proceed we need a conformal writing of the metric. For this let us recall that ${\mb H}^2$ is embedded into ${\mb R}^3$ according to
\beq
x_1^2+x_2^2-x_3^2=-1,\qq (x_1,x_2)\in\,{\mb R}^2,\qq x_3\geq 1.\eeq
The coordinates choice
\beq
x_1=\cosh\chi\sinh y,\qq x_2=\sinh\chi,\qq x_3=\cosh\chi\cosh y,\qq (\chi,y)\in\,{\mb R}^2\eeq
gives
\beq
g_0({\mb H}^2)\equiv dx_1^2+dx_2^2-dx_3^2=d\chi^2+\cosh^2\chi\,dy^2.\eeq

So we can define the conformal factor $\rho$ by
\beq\label{globM}
A^2\,dt^2+\cosh^2 t\,dy^2=\rho^2(d\chi^2+\cosh^2\chi\,dy^2)=\rho^2\,g_0({\mb H}^2),\eeq
as well as  
\beq
\psi(t)\equiv\sum_{k=1}^{\nu}\,\arctan\big(h_k(t)\big),\qq h_k(t)=e_k\sqrt{m_k\cosh^2 t-1}, \eeq
and
\beq\Si^{(\nu)}(t)\equiv \cos\psi(t)-\sin\psi(t)\,\sinh t.\eeq
Let us prove
\begin{nlem}\label{var} If $\Si^{(\nu)}>0$  then the manifold on which the metric (\ref{globM}) is defined will be diffeomorphic to ${\mb H}^2$. A zero of 
$\Si^{(\nu)}$ precludes any manifold.
\end{nlem}

\vspace{3mm}
\nin{\bf Proof:}  The coordinate $\chi$ and the conformal factor $\rho$ are given by
\beq\label{eqconf}
\rho\,D_t\chi=A, \qq \& \qq \rho\,\cosh\chi=\cosh t.
\eeq
Dividing these two relations and integrating gives
\beq\label{chi}
e^{\chi}=
\frac{e^t+\tan(\psi(t)/2)}{1-e^t\,\tan(\psi(t)/2)}=\frac{e^t\,\Si^{(\nu)}(t)}{(\cos(\psi(t)/2)-\sin(\psi(t)/2)\,\sinh t)^2},\eeq
The strict positivity of 
$e^{\chi}$ is therefore equivalent to $\Si^{(\nu)}>0$.

Computing $D_t\chi$ and plugging it into the first relation in (\ref{chi}) gives $\rho=\Si^{(\nu)}$.

The relation (\ref{chi}) allows to compute $\cosh\chi$ and to check that the second relation in (\ref{eqconf}) gives anew $\rho=\Si^{(\nu)}$.

So, if the conformal factor $\rho=\Si^{(\nu)}(t)$ {\em never} vanishes then the manifold is $M\cong{\mb H}^2$. 
$\hfill\Box$

\subsection{Proof of point b) in Theorem 5}
The case of integrals quadratic in the momenta,   
due to Koenigs, was already settled in Section 6. In this case the manifold is $M\cong {\mb H}^2$.

Let us consider now the case  of higher integrals in the momenta with $\sharp(S_i)=2n$ for $n\geq 2$ in which case  we have $\nu=2n-1$. As proved in Appendix B we have
\beq
\Si^{(2n-1)}(t)=\frac 1{{\cal S}_{2n-1}(\cosh t)^{2n-1}}\Big(\sum_{l=0}^{n-1}(-1)^lH_{2l}^{(2n-1)}(t)-\sinh t\,\sum_{l=0}^{n-1}(-1)^lH_{2l+1}^{(2n-1)}\Big),\eeq 
where
\beq
\prod_{k=1}^{2n-1}(1+\xi\,\sqrt{m_k})=\sum_{k=0}^{2n-1}\xi^k{\cal S}_k,\qq \Longrightarrow \quad {\cal S}_{2n-1}=\prod_{k=1}^{2n-1}\sqrt{m_k}.\eeq
For $t\to \pm\nf$ we have the equivalents
\beq
\frac 1{{\cal S}_{2n-1}(\cosh t)^{2n-1}}\sum_{l=0}^{n-1}(-1)^lH_{2l}^{(2n-1)}(t)\sim (-1)^{n-1}\frac{A(+\nf)}{\cosh t}
\eeq
and
\beq
-\frac{\sinh t}{{\cal S}_{2n-1}(\cosh t)^{2n-1}}\sum_{l=0}^{n-1}(-1)^lH_{2l+1}^{(2n-1)}\sim (-1)^n\,\sinh t.\eeq
Hence $\Si^{(2n-1)}(\pm\nf)$ are of opposite signs and this implies a zero of $\Si^{(2n-1)}(t)$, precluding any manifold.$\hfill\Box$

%%%%%%%%%%%%%%%%%%%%%%%%%%%%%%%%%%%%%%%%%%
\nc{\wwh}{\wti{h}} \nc{\wwm}{\wti{m}}  
\nc{\wmu}{\wti{\mu}}
%%%%%%%%%%%%%%%%%%%%%%%%%%%%%%%%%%%%%%%%%

\subsection{Towards the proof of point c) in Theorem 5}
Here we need some preparatory material. First we have 
$\sharp(S_i)=2n+1$, with $n\geq 1$, and so $\nu=2n$.

We have $2n$ ``masses"  
\[ \{m_1,\ldots,m_n,\wwm_1,\ldots,\wwm_n\},\qq e_{n+k}=-e_k,\quad \forall k\in\{1,\ldots,n\}.\]
This just means that we take $\dst\sum_{k=1}^{2n}\,e_k=0$. These choices give 
\beq
A^{(2n)}(t)=1+{\cal A}^{(2n)}(t),\qq {\cal A}^{(2n)}(t)=\sum_{k=1}^n\,\sinh t\left(\frac 1{h_k(t)}-\frac 1{\wwh_k(t)}\right),\eeq
where
\beq\label{defh}
h_k(t)=e_k\sqrt{m_k\cosh^2 t-1},\qq\qq \wwh_k(t)=e_k\sqrt{\wwm_k\cosh^2 t-1},\eeq
as well as
\beq
\psi^{(2n)}(t)=\sum_{k=1}^n\Big(\arctan(h_k)-\arctan(\wwh_k)\Big),\eeq
and
\beq\label{defHH}
\prod_{k=1}^n(1+\xi\,h_k(t))(1-\xi\,\wwh_k(t))=\sum_{l=0}^{2n}\,H^{2n}_l(t)\,\xi^n.\eeq

Let us begin with
\begin{nlem}\label{condition} The inequality
\beq
\sum_{k=1}^n\left|\frac 1{\sqrt{\mu_k}}-\frac 1{\sqrt{\wmu_k}}\right|<1,\qq\mu_k=m_k-1,\quad \wmu_k=\wwm_k-1,\eeq
implies that
\beq
\Big(\forall t\in{\mb R}:\qq A^{(2n)}(t)>0\Big)\qq\&\qq  
\ A^{(2n)}(\pm\nf)>0.\eeq
\end{nlem}

\vspace{3mm}
\nin{\bf Proof:} Defining $\tau=\tanh t$ and $\mu_k=m_k-1,\  \wmu_k=\wwm_k-1$ gives
\beq
{\cal A}^{(2n)}=\sum_{k=1}^n\left(\frac{\tau}{\mu_k+\tau^2}-\frac{\tau}{\wmu_k+\tau^2}\right), \qq \tau\in\,[-1,+1].\eeq
So we have
\beq
|{\cal A}^{(2n)}|\leq \sum_{k=1}^n\frac{|\mu_k-\wmu_k|}{\sqrt{\mu_k+\tau^2}\sqrt{\wmu_k+\tau^2}(\sqrt{\mu_k+\tau^2}+\sqrt{\wmu_k+\tau^2})},
\eeq
from which we deduce
\beq
|{\cal A}^{(2n)}|\leq \sum_{k=1}^n\frac{|\mu_k-\wmu_k|}{\sqrt{\mu_k}\sqrt{\wmu_k}(\sqrt{\mu_k}+\sqrt{\wmu_k})}\leq \sum_{k=1}^n\frac{|\sqrt{\mu_k}-\sqrt{\wmu_k}|}{\sqrt{\mu_k}\sqrt{\wmu_k}}<1\qq \forall \tau\in\,[-1,+1].\eeq
It follows that $A(t)$ will be strictly positive not only for $t\in{\mb R}$ but also for $t\to\,\pm\nf$. $\hfill\Box$ 

The next step is
\begin{nlem}\label{recH} One has the following relations:
\beq\barr{clcc}
l=0: & H_0^{2n}=H_0^{2(n-1)} &  &   \\ [4mm]
l=1: & H_1^{2n}=H_1^{2(n-1)} & +(h_n-\wwh_n)\,H_0^{2(n-1)} &   \\[4mm]
2\leq l \leq 2(n-1): & H_l^{2n}=H_l^{2(n-1)} & +(h_n-\wwh_n)H_{l-1}^{2(n-1)} & -h_n\wwh_n\,H_{l-2}^{2(n-1)}\\[4mm]
l=2n-1: & H_{2n-1}^{2n}= & +(h_n-\wwh_n)H_{2(n-1)}^{2(n-1)} & -h_n\wwh_n\,H_{2n-3}^{2(n-1)}\\[4mm]
l=2n: & H_{2n}^{2n}= &  &  -h_n\wwh_n\,H_{2(n-1)}^{2(n-1)}.\earr\eeq
\end{nlem}

\vspace{3mm}
\nin{\bf Proof:} Relation (\ref{defHH}) implies
\beq
(1+\xi\,h_n)(1-\xi\,\wwh_n)\sum_{l=0}^{2(n-1)}\,H^{2(n-1)}_l(t)\,\xi^l=\sum_{l=0}^{2n}\,H^{2n}_l(t)\,\xi^l.
\eeq
Expanding both sides in powers of $\xi$ gives (\ref{recH}).$\hfill\Box$

In Appendix B it is proved that     
\beq\label{cos}
\cos\psi^{(2n)}(t)=\frac 1{{\cal S}_n(\cosh t)^{2n}}\sum_{l=0}^n(-1)^l\,H_{2k}^{2n}(t),\eeq
and
\beq\label{sin}
\sin\psi^{(2n)}(t)=\frac 1{{\cal S}_n(\cosh t)^{2n}}\sum_{l=0}^{n-1}(-1)^l\,H_{2k+1}^{2n}(t),\qq {\cal S}_n=\sqrt{\prod_{k=1}^n\,m_k\,\wwm_k}.     
\eeq     
From these relations we deduce 
\begin{nlem}\label{sincos} One has the following recurrences
\beq
\cos\psi^{(2n)}=\frac 1{\sqrt{m_n\wwm_n}}\left(\frac{(1+h_n\wwh_n)}{\cosh^2 t}\,\cos\psi^{(2n-2)}-\frac{(h_n-\wwh_n)}{\cosh^2 t}\,\sin\psi^{(2n-2)}\right),
\eeq
and
\beq
\sin\psi^{(2n)}=\frac 1{\sqrt{m_n\wwm_n}}\left(\frac{(1+h_n\wwh_n)}{\cosh^2 t}\,\sin\psi^{(2n-2)}+\frac{(h_n-\wwh_n)}{\cosh^2 t}\,\cos\psi^{(2n-2)}\right).
\eeq\end{nlem}

\vspace{3mm}
\nin{\bf Proof:} Use relations (\ref{cos}) and (\ref{sin}) and Lemma \ref{recH}.$\hfill\Box$

Similarly we have
\begin{nlem}\label{recSi} Having defined
\beq
\Si^{(2n)}=\cos\psi^{(2n)}-\sinh t\,\sin\psi^{(2n)},
\eeq
imples the recurrence relation
\beq
\Si^{(2n)}=\frac 1{\sqrt{m_n\wwm_n}}\left(\left[1+\frac{(h_n+\sinh t)(\wwh_n-\sinh t)}{\cosh^2 t}\right]\,\Si^{(2n-2)}+(\wwh_n-h_n)\,\sin\psi^{(2n-2)}\right).\eeq
\end{nlem}

\vspace{3mm}
\nin{\bf Proof:} Use relations (\ref{cos}) and (\ref{sin}) and Lemma \ref{sincos}.$\hfill\Box$

To prepare the final proof we need
\begin{nlem}\label{init} For $n=1$ we have for manifold $M\cong{\mb H}^2$.
\end{nlem}

\vspace{3mm}
\nin{\bf Proof:} We have
\beq\barr{l}\dst 
\Si^{(2)}(t)=\frac{H_0^{2}-H_2^{2}-\sinh t\,H_1^{2}}{\sqrt{m_1\wwm_1}\cosh^2 t}=\frac{1+h_1\wwh_1-(h_1-\wwh_1)\sinh t}{\sqrt{m_1\wwm_1}\cosh^2 t}= \\[4mm]\dst 
\hspace{3cm}=\frac{(h_1+\sinh t)(\wwh_1-\sinh t)+\cosh^2 t}{\sqrt{m_1\wwm_1}\cosh^2 t}>0 \qq\forall t\in\,{\mb R}.\earr\eeq

Since strict inequalities are not respected when taking limits, we must check that $\Si^{(2)}(\pm\nf)>0$. This follows from
\beq
\Si^{(2)}(\pm\nf)=1+\frac 1{\sqrt{m_1}}-\frac 1{\sqrt{\wwm_1}}.
\eeq
Since $m_1>1$ and $\wwm_1>1$, we have
\beq
-1<\frac 1{\sqrt{m_1}}-\frac 1{\sqrt{\wwm_1}}<+1\qq\Longrightarrow\qq \Si^{(2)}(\pm\nf)>0.\eeq
Use of Lemma \ref{var} concludes the proof.$\hfill\Box$ 

Before giving the proof of point c) in Theorem 5, let us give the precise hypotheses needed:
\brm
\item[\bf (h1)] In relation (\ref{defh}) we take all $e_k=1$. 
\item[\bf (h2)] For the ``masses" we have
\[\forall k\in\,\{1,\ldots,n-1\}:\quad m_k>\wwm_k>1,\qq\quad 1<m_n<\wwm_n,\]
\item[\bf (h3)] And the bound
\[\sum_{k=1}^n\left|\frac 1{\sqrt{\mu_k}}-\frac 1{\sqrt{\wmu_k}}\right|<1,\qq \mu_k=m_k-1,\quad \wmu_k=\wwm_k-1.\] 
\erm
Under these hypotheses let us prove:

\subsection{Proof of point c) in Theorem 5}
For $n=1$ we have proved in Lemma \ref{init} that 
$\Si^{(2)}$ never vanishes. Let us proceed by recurrence. Lemma \ref{recSi} gives
\beq
\Si^{(2n)}=\frac 1{\sqrt{m_n\wwm_n}}\left(\left[1+\frac{(h_n+\sinh t)(\wwh_n-\sinh t)}{\cosh^2 t}\right]\,\Si^{(2n-2)}+(\wwh_n-h_n)\,\sin\psi^{(2n-2)}\right).
\eeq
Hypothesis ${\bf h2}$ implies that $\wwh_n-h_n$ is strictly positive, so if $\sin\psi^{(2n-2)}>0$ we can  conclude that $\forall t\in\,{\mb R}$ we have $\Si^{(2n)}>0$.

Let us recall that
\beq
\psi^{(2n-2)}(t)=\sum_{k=0}^{n-1}(\arctan h_k(t)-\arctan \wwh_k(t))>0,
\eeq
with 
\beq
D_t\psi^{(2n-2)}=\frac{A^{(2n-2)}-1}{\cosh t}, \qq\quad A^{(2n-2)}-1=\sum_{k=1}^{n-1}\sinh t\left(\frac 1{h_k}-\frac 1{\wwh_k}\right).\eeq 
Using these relations one can easily establish the bounds
\beq
0<\psi^{(2n-2)}(t)\leq B_n\equiv\sum_{k=0}^{n-1}(\arctan \sqrt{\mu_k}-\arctan \sqrt{\wmu_k}).\eeq
The upper bound becomes
\beq
B_n=\sum_{k=0}^{n-1}\left(\arctan\frac 1{\sqrt{\wmu_k}}-\arctan\frac 1{\sqrt{\mu_k}}\right)=\sum_{k=0}^{n-1}\arctan\left(\frac{\frac 1{\sqrt{\wmu_k}}-\frac 1{\sqrt{\mu_k}}}{1+\frac 1{\sqrt{\mu_k\,\wmu_k}}}\right)\eeq
leading to
\beq
B_n<\sum_{k=0}^{n-1}\arctan\left(\frac 1{\sqrt{\wmu_k}}-\frac 1{\sqrt{\mu_k}}\right)\leq\sum_{k=0}^{n-1}\left|\frac 1{\sqrt{\wmu_k}}-\frac 1{\sqrt{\mu_k}}\right|<\sum_{k=0}^n\left|\frac 1{\sqrt{\wmu_k}}-\frac 1{\sqrt{\mu_k}}\right|.\eeq
So we get $0<\psi_0^{(2n-2)}<1<\pi/2$ implying the strict  positivity of $\sin\psi^{(2n-2)}(t)$.

Strict inequalities are not respected when taking limits, so we have to check that $\Si^{(2n)}(\pm\nf)>0$. Since we have
\beq
\Si^{(2n)}(\pm \nf)=A^{(2n)}(\pm\nf),\eeq
the hypothesis ${\bf (h3)}$ and Lemma \ref{condition} conclude the proof.$\hfill\Box$

This ends up the proof of point c) in Theorem 5.$\hfill\Box$

\section{Conclusion}
Let us add the following remarks:
\brm
\item Once more let us point out that SI geodesic flows are not necessarily related to Zoll geometry. Conversely it is not known whether any Zoll metric of revolution, globally defined on 
${\mb S}^2$, produces a SI geodesic flow. 
\item We have proved the {\em existence} of a solution for the differential systems in Propositions 1 and 10. However the problem of {\em uniqueness} is left open.
\item Let us recall an important open problem: do there exist SI geodesic flows beyond the hypotheses of Matveev and Shechishin?
\item Locally there is an intriguing ``symmetry" between the trigonometric and the hyperbolic cases but {\em globally} they are drastically different: 
\begin{itemize}
\item In the trigonometric case we may find metrics defined on ${\mb S}^2$,
\item In the hyperbolic case we may find metrics defined on ${\mb H}^2$!
\end{itemize} 
\item It is interesting to compare the two cases:
\[\barr{lcc}
\hspace{2cm} &\fbox{trigonometric}  & \fbox{hyperbolic}\\[4mm]
\hspace{2cm} & g=A^2(t)dt^2+\sin^2 t\,dy^2 & g=A^2(t)dt^2+\cosh^2 t\,dy^2\\[4mm] 
\hspace{2cm} & t\in\,(0,\pi),\ y\in\,{\mb S}^1 & (t,y)\in\,{\mb R}^2\\[4mm]
\hspace{2cm} & \dst A(t)=1+\sum_{k}\frac{e_k\,\sin t}{\sqrt{1-m_k\,\sin^2 t}}  & \dst A(t)=1+\sum_{k}\frac{e_k\,\sinh t}{\sqrt{m_k\,\cosh^2 t-1}}\\ [6mm]
\sharp(S)=2 \ (\mbox{Koenigs}): &   \mbox{no manifold} &  {\mb H}^2\\[4mm]
\sharp(S)=2n\geq 4: &   
\mbox{no manifold} & \mbox{no manifold}\\ [4mm] 
\sharp(S)=2n+1\geq 3: & \mbox{Zoll}\ \to\ {\mb S}^2  &  {\mb H}^2\earr\]
\item Leaving aside Koenigs system, there seems to be a curse in the case $\sharp(S)=2n$ for which no manifold seems to be allowed. There should be an explanation of this fact.
\erm

\begin{appendices}

\section{Appendix A}
Let us recall the content of Definition \ref{newH}. We have 
\[\forall k\in\{1,2,\ldots,\nu\}: \quad h_k(t)=e_k\,\sqrt{m_k\,\cosh^2 t-1}\qq \forall k:\ \Big( e_k=\pm 1\quad \&\quad  m_k\geq 1\Big),\]
as well as
\beq A(t)=1+\sum_{k=1}^{\nu}\frac{\sinh t}{h_k(t)}.\eeq

The functions $H_k^{\nu}(t)$ are defined by the generating function
\beq\label{AppdefH}
{\cal H}^{\nu}(t,\xi)\equiv\prod_{k=1}^{\nu}(1+\xi\,h_k(t))=\sum_{k=0}^{\nu}\,H_k^{\nu}(t)\,\xi^k.\eeq
For convenience we will introduce the conventional values $H_{-1}^{\nu}(t)=H_{-2}^{\nu}(t)=H^{\nu}_{\nu+1}(t)\equiv 0$.

Let us prove some useful relations:

\begin{nth} For all $k\in\{0,1,\ldots,\nu\}$ one has
\beq\label{derH}
D_t\,H_{k}^{\nu}(t)=\tanh t\Big(k\,H_k^{\nu}+(k-\nu-2)\,H_{k-2}^{\nu}\Big)+\frac{(A-1)}{\cosh t}H_{k-1}^{\nu},
\eeq
as well as
\beq\label{Hspe}
\sinh t\ H_{\nu-1}^{\nu}=(A-1)\,H_{\nu}^{\nu}.\eeq\end{nth}

\vspace{5mm}
\nin{\bf Proof:} The relation is trivial for $k=0$, so let it be supposed that $k\geq 1$. Using the relations
\beq
h'_k=\tanh t\left(h_k+\frac 1{h_k}\right),
\eeq
we deduce
\beq
\frac{\pt_t{\cal H}}{\cal H} =\tanh t\,\left(\sum_{k=1}^{\nu}\frac{\xi h_k}{1+\xi h_k}+\sum_{k=1}^{\nu}\frac \xi{h_k(1+\xi h_k)}\right).
\eeq
The first sum is merely
\beq
\sum_{k=1}^{\nu}\frac{\xi h_k}{1+\xi h_k}= \frac{\xi\pt_{\xi}{\cal H}}{\cal H},\eeq
while the second sum is transformed according to
\beq\barr{lcl}\dst 
\sum_{k=1}^{\nu}\frac{\xi(1-\xi^2 h_k^2+\xi^2 h_k^2)}{h_k(1+\xi h_k)} & = & \dst \sum_{k=1}^{\nu}\frac{\xi(1-\xi h_k)}{h_k}+\xi^2\sum_{k=1}^{\nu}\frac{\xi h_k}{1+\xi h_k}\\[5mm] & = & \dst \xi\frac{(A-1)}{\sinh t}-\nu\,\xi^2+\xi^2\,\frac{\xi\pt_{\xi}{\cal H}}{\cal H}.\earr
\eeq
Hence we have obtained
\beq
\pt_t{\cal H}=\tanh t\Big((1+\xi^2)\xi\pt_{\xi}{\cal H}-\nu\xi^2\,{\cal H}\Big)+\xi\frac{(A-1)}{\cosh t}\,{\cal H}.
\eeq
Expanding in powers of $\xi$ gives (\ref{derH}) for $k$ 
from $1$ to $\nu$. For $k=\nu+1$ we get (\ref{Hspe}) 
while for $k=\nu+2$ the relation obtained is trivial.
$\hfill\Box$

We will need also, for the functions
\beq
{\cal H}_{\pm}^{\nu}={\cal H}^{\nu}(t,\pm i\eta),\eeq
the following:

\begin{nth} For $\nu=2n$ one has the relations
\beq\label{Hpm}
\frac{{\cal H}_+^{2n} +{\cal H}_-^{2n}}{2}=\sum_{l=0}^{n}(-1)^l\,H_{2l}^{2n}\,\eta^{2l},\qq 
\frac{{\cal H}_+^{2n} -{\cal H}_-^{2n}}{2i}=\sum_{l=0}^{n-1}(-1)^l\,H_{2l+1}^{2n}\,\eta^{2l+1}.
\eeq
\end{nth}

\vspace{5mm}
\nin{\bf Proof:} Starting from (\ref{AppdefH}) with $\xi\ \to\ \pm\,i\eta$, expanding in powers of $i\eta$ and separating the real and imaginary parts proves both  relations.$\hfill\Box$ 

\section{Appendix B}
Let us come back to relation (\ref{AppdefH}) for $\xi=i$. We have
\beq
\prod_{k=1}^{\nu}(1+ih_k(t))=\sum_{k=0}^{\nu}\,i^kH_k^{\nu}(t).\eeq
Noticing that $1+ih_k(t)=\sqrt{m_k}\cosh t\,e^{i\arctan h_k(t)}$ the previous relation becomes
\beq
\prod_{k=1}^{\nu}\sqrt{m_k}\cosh^{\nu} t\,e^{i\psi(t)}=\sum_{k=0}^{\nu}\,i^kH_k^{\nu}(t),\qq \psi(t)=\sum_{k=1}^{\nu}\arctan\big(h_k(t)\big).
\eeq
Defining $\dst {\cal S}_{\nu}=\prod_{k=1}^{\nu}\sqrt{m_k}$ and  
comparing the real and the imaginary gives $\cos\psi(t)$ and $\sin\psi(t)$ which lead, for $\nu=2n-1$, to
\beq
\Si^{(2n-1)}(t)=\frac 1{{\cal S}_{2n-1}(\cosh t)^{2n-1}}\Big(\sum_{l=0}^{n-1}(-1)^lH_{2l}^{2n-1}(t)-\sinh t\,\sum_{l=0}^{n-1}(-1)^lH_{2l+1}^{2n-1}\Big).\eeq 
For $\nu=2n$ we get similarly the relation 
\beq
\Si^{(2n)}(t)=\frac 1{{\cal S}_{2n}(\cosh t)^{2n}}\Big(\sum_{l=0}^{n}(-1)^l\,H_{2l}^{2n}(t)-\sinh t\,\sum_{l=0}^{n-1}(-1)^l\,H_{2l+1}^{2n}\Big).
\eeq

\end{appendices}

\end{document}